\documentclass[preprint] {article}
\usepackage{amsmath}
\usepackage{amsfonts}
\usepackage{amssymb}
\usepackage{geometry}
\usepackage{graphicx}
\usepackage{braket}
\allowdisplaybreaks
\newcommand {\bp}{\begin{pmatrix}}
\newcommand {\ep}{\end{pmatrix}}
\newcommand{\be}{\begin{equation}} \newcommand{\ee}{\end{equation}}
\newcommand{\bea}{\begin{eqnarray}}\newcommand{\eea}{\end{eqnarray}}

\begin{document}
\title{Integrable coupled Li$\acute{e}$nard-type systems with 
balanced loss and gain }

\author{ 
Debdeep Sinha\footnote{{\bf email:}  debdeepsinha.rs@visva-bharati.ac.in}\ and
Pijush K. Ghosh\footnote {{\bf email:} 
pijushkanti.ghosh@visva-bharati.ac.in}} 
\date{Department of Physics, Siksha-Bhavana, \\ 
Visva-Bharati University, \\
Santiniketan, PIN 731 235, India.}
\maketitle

\begin{abstract}

A Hamiltonian formulation of generic many-particle systems with 
space-dependent balanced loss and gain coefficients is presented.
It is shown that the balancing of loss and gain necessarily occurs
in a pair-wise fashion. Further, using a suitable choice of co-ordinates,
the Hamiltonian can always be reformulated as a many-particle system
in the background of a pseudo-Euclidean metric and subjected to an analogous
inhomogeneous magnetic field with a functional form that is identical with
space-dependent loss/gain co-efficient.The resulting equations of motion
from the Hamiltonian are a system of coupled 
Li$\acute{e}$nard-type differential equations.  Partially integrable
systems are obtained for two distinct cases, namely, systems with (i)
translational symmetry or (ii) rotational invariance in a pseudo-Euclidean
space. A total number of $m+1$ integrals of motion are constructed for a
system of $2m$ particles, which are in involution, implying that two-particle
systems are completely integrable. A few exact solutions for both the cases
are presented for specific choices of the potential and space-dependent
gain/loss co-efficients, which include periodic stable solutions. Quantization
of the system is discussed with the construction of the integrals of
motion for specific choices of the potential and gain-loss coefficients. A few
quasi-exactly solvable models admitting bound states in appropriate Stoke
wedges are presented.
\end{abstract}

\tableofcontents
\vspace{0.3in}

\section{Introduction}

Dissipative systems are ubiquitous in nature. One of the approaches of having a
Hamiltonian formulation for dissipative harmonic oscillator is due to
Bateman \cite{bat}, in which an auxiliary system is introduced as a thermal
bath that is time-reversed version of the original oscillator. The dissipative 
oscillator and its auxiliary system taken together give rise to a Hamiltonian
with equally balanced loss and gain terms. Various issues related to the
quantization of Bateman-type of oscillators are discussed in
Refs.\cite{bopp,fes,trikochinsky, dekker, rasetti,rabin,jur}.

With the technological advancements, tailoring systems with balanced loss and
gain is a reality\cite{bpeng,ben}. One of the important features of this system
is the existence of stable bound states within certain regions of
parameter-space, when the system and bath are suitably coupled\cite{ben}. In
order to explore this class of systems further, the Hamiltonian formulation of
generic
many-particle systems with balanced loss and gain in a model independent manner
is required. Apart from being an important ingredient in the investigations of
systems with balanced loss and gain, such a formulation may also be used
to study the purely dissipative dynamics by exploiting tools and techniques
associated with a Hamiltonian system. It may be noted that until recently there
were a very few examples of systems with balanced loss and gain for which
Hamiltonian formulations were available\cite{ben1,ivb,sagar,ds-pkg,khare}. Further,
such constructions were specific to the model under investigations. Within this
background, the Hamiltonian formulation of a generic many-body system with
balanced loss and gain is presented in a systematic way in Ref. \cite{pkg-ds}.
It is shown that the Hamiltonian formulation is possible only if the balancing
of loss and gain occurs in a pair-wise fashion. It is also shown that with a
choice of a suitable coordinate the Hamiltonian can always be formulated as
describing a many-particle system in the background of a pseudo-Euclidean
metric and subjected to an external analogous uniform magnetic field. A few
exactly solvable models are
presented with the construction of a set of integrals of motion.  
The quantization of the exactly solvable models presented in Ref. \cite{pkg-ds}
is considered in Ref. \cite{ds-pkg1} with a construction of the many-body
correlation functions of a class of Calogero-type models with balanced loss
and gain by mapping the relevant integrals to the known results of
random matrix theory.

The Li$\acute{e}$nard  equation\cite{lie,van,hn,guha} exhibits many novel
mathematical features such as limit cycles, isochronicity, etc.
and finds widespread applications in many branches of applied sciences.
The Van der Pol oscillator \cite{van}, which is a particular form of
Li$\acute{e}$nard equation, also perceive plenty of applications in physical
\cite{p,p1, p2}, chemical\cite{c}, biological \cite{b} and mathematical
\cite{m} sciences. A characterizing feature of Li$\acute{e}$nard equation is
that the dissipative term is space-dependent. Consequently, depending on the
specific form of the space-dependent coefficient of the term linear in
velocity, the system may have gain in some regions of space and loss elsewhere.
One important aspect of this space-dependent gain-loss term is the existence
of limit cycles and relaxation oscillation.

The main purpose of the present article is to consider the Hamiltonian
formulation presented in Ref. \cite{pkg-ds} and to extend it to the case
when the loss-gain coefficients are space dependent. In particular,
the Hamiltonian formulation of many-particle systems in presence of
space-dependent balanced loss and gain coefficients is presented in a model
independent way. The generic features of Hamiltonian systems with constant
coefficients for the balanced loss and gain terms persist even if these
coefficients are allowed to be space-dependent. In particular, with appropriate
choice of the co-ordinates, the Hamiltonian with space-dependent loss and gain
can always be reformulated as a many-particle system in the background
of a pseudo-Euclidean metric and subjected to an analogous external
inhomogeneous magnetic field having the functional form same as space-dependent
balanced loss/gain coefficients.
Further, the balancing of space-dependent loss and gain terms necessarily occurs
in a pair-wise fashion. The resulting equations of motions from the Hamiltonian
are coupled Li$\acute{e}$nard-type of differential equations. A region of gain
for a particle is a region of loss for the corresponding paired particle and
the vice verse. This raises the possibility of existence of stable bound states
within a certain region of parameter-space, even if neither a particle nor
its paired particle admits limit cycles.

The Hamiltonian formulation of systems with balanced loss and gain can also
be used to investigate purely dissipative dynamics by choosing the many-body
potential judiciously such that only `unidirectional coupling' is allowed. In
particular, the dynamics of the dissipative system is made independent of the
dynamics of its auxiliary system. However, the dynamics of the auxiliary
system is dependent on the dynamics of the dissipative system, thereby allowing
only `unidirectional coupling'. The dissipative and the corresponding auxiliary
systems taken together are described by a Hamiltonian. The advantage of such a
construction is that techniques associated with Hamiltonian formulation like
canonical perturbation theory, canonical quantization, KAM theory etc. may be
used to study purely dissipative dynamics. A generic Hamiltonian
formulation of systems with balanced loss and gain and with unidirectional
coupling is presented in this article. The examples considered are dissipative
rational as well as trigonometric Calogero-Sutherland models associated with
various root structures and dissipative Toda system.

The integrability and exact solvability of Hamiltonian systems with balanced
loss and gain are investigated when the potential admits a translational
symmetry or a rotational symmetry in a pseudo-Euclidean space. A set of $m+1$
integrals of motion is obtained for both the cases for a system of $N=2m$
particles. These integrals of motion are in involution, implying that the
system is partially integrable for $N > 2$, while it is completely integrable
for $N=2$. A few exact analytical solutions are obtained for specific choices
of the potential and the space-dependent loss/gain profile for translational
as well rotationally invariant systems. Stable bound states exist within
certain ranges in the parameter-space. Quantization of the system is carried out
with the construction of the integrals of motion for specific choices of
the potential and gain-loss coefficients. For the quantum case, normalizable
solutions are obtained for a few quasi-exactly solvable models.

The paper is organized in the following manner. In the next section the
Lagrangian and Hamiltonian formulation for many-body systems with space
dependent balanced loss and gain coefficients is presented in a  model
independent way and the equations of motion are obtained. It is shown that
the system can always be reformulated as a many-particle system in the back
ground of a pseudo-Euclidean metric by using a suitable choice of the
co-ordinates. Section-3 deals with the space dependent balanced loss and gain
systems, when the system admits a translational symmetry. In section-4, the
case for rotationally symmetric system is considered in a pseudo-Euclidean
space. In section-5, the quantization of the classical 
Hamiltonian is carried out with the construction of the integrals of motion.
For the quantum case, normalizable solutions are presented for some of the
quasi-exactly solvable models. Finally, in the last section, the results are 
summarized with a discussion. Some of the equivalent Lagrangian for many-body
systems with space dependent balanced loss and gain coefficients are presented 
in the Appendix-A.

\section{Hamiltonian formulation}

A Hamiltonian formulation of many-particle systems with balanced
loss and gain is presented in Ref. \cite{pkg-ds}. The loss/gain coefficient
is constant in this approach and the Hamiltonian is written as,
\bea
H=\Pi^T M \Pi + V(x_1, x_2, \dots x_N),
\label{H}
\eea 
\noindent where $M$ is a $N \times N$ real symmetric matrix with 
$X=(x_1, x_2, \dots x_N)^T$ and $\Pi=(\pi_1, \pi_2, \dots \pi_N)^T$
are $N$ coordinates and their conjugate momenta, respectively. The suffix $T$
in $O^T$ denotes the transpose of a matrix $O$. The generalized momenta
is defined by
$\Pi=P + A X$, where $A$ is an $N \times N$ anti-symmetric matrix. This
analysis excludes constrained systems and any non-standard Hamiltonian
formulation. Systems with the dissipative term depending nonlinearly on
the velocity are not under the purview of the present investigation. A suggestion
is made in Ref. \cite{pkg-ds} that this Hamiltonian
formulation can also be generalized to include space-dependent balanced loss
and gain co-efficients by redefining the generalized momenta $\Pi$ as,
\bea
\Pi=P+A F(X),
\label{gm}
\eea
where $F(X)=(F_1, F_2, \dots F_N)^T$ is $N$ dimensional column matrix whose
entries are functions of coordinates. The choice $F_i\equiv \frac{x_i}{2}$
corresponds to systems with constant loss/gain co-efficients. 
This scheme for space-dependent balanced loss/gain was implemented\footnote{
It may be noted that Eq.(16)  in Ref.  \cite{pkg-ds} contains a
typographical error. It is valid for $N=2$ instead of arbitrary $N$.} for
$N=2$ and as an example, Hamiltonian for the Van der pol oscillator with
balanced loss and gain was constructed with the choice $F_i\equiv\frac{1}{2} x_i
-\frac{1}{3} x_i^3$. The analysis for $N >2$ is much more involved and
needs separate investigations.

In the present article, a Hamiltonian formulation of many-particle systems
with space dependent balanced loss and gain coefficients is presented for
arbitrary $N$. The equations of motion derived from the Hamiltonian (\ref{H})
with the generalized momenta $\Pi$ defined by Eq. (\ref{gm}) has the following
form:
\bea
&&\ddot{X}-2MR\dot{X}+2M\frac{\partial V}{\partial X}=0,\nonumber \\
&& [J]_{ij} \equiv \frac{\partial F_i}{\partial x_j},\ \
R \equiv AJ-(AJ)^T, \ \  \frac{\partial V}{\partial X}\equiv
\left(\frac{\partial V}{\partial x_i}, \frac{\partial V}{\partial x_2},
\dots \frac{\partial V}{\partial x_N}\right)^T.
\label{X}
\eea
%The resulting equations of motions are a system of coupled
% Li$\acute{e}$nard-type differential equations with balanced loss and gain.  
The above equations can also be derived from the following Lagrangian:
\be
{\cal{L}} = \frac{1}{4}\dot{X}^T M^{-1} \dot{X} -\frac{1}{2} (\dot{X}^T AF +F^T A^T \dot{X}) -
V(x_1, x_2, \dots, x_N).
\label{lag}
\ee
\noindent 
A phase-space analysis of Eq. (\ref{X}) shows that only the diagonal elements
of the matrix $M R$ are relevant for determining whether or not the system
is dissipative. The following condition is imposed by demanding that the
velocity dependent term in the equation of motion for $x_i$ should only contain
$\dot{x}_i$ :
\bea
MR= D,
\label{1con}
\eea
where $D$ is a diagonal matrix. It may be noted that unlike the case of
constant gain/loss coefficients\cite{pkg-ds}, both $D$ and $R$ for the present
case depend on spatial variables. Nevertheless, it can be shown by using
Eq. (\ref{1con}) and the properties of $M$, $R$, $D$, i.e. $M^T=M, D^T=D,
R^T=-R$ that,
\be
\{M, R\}=0, \ \{M, D\}=0, \ \{R, D\}=0,
\label{anti}
\ee 
\noindent where $\{,\}$ denotes anticommutaror. It immediately follows that $Tr(D)=0$, implying that gain and loss
are equally balanced. A general discussion on the properties of the matrices
$M, R, D$ and its consequences on the physical behavior of the system is
given in Ref. \cite{pkg-ds}, which are equally applicable for the case
of space-dependent loss/gain coefficients for which $D$ and $R$ are also
space-dependent. However, the representations of these matrices are different
for these two cases.

\noindent As in the case of systems with constant loss/gain co-efficients
\cite{pkg-ds}, the Hamiltonian of Eq. (\ref{H}) with $\Pi$ given in Eq.
(\ref{gm}) may be re-interpreted as defined in the background of a
pseudo-Euclidean metric. The symmetric matrix $M$ can always be diagonalized
to $M_d$ by using an orthogonal matrix $\hat{O}$, i.e. $M_d=\hat{O}M\hat{O}^T$.
The anti-commuting property of $M$ with $D$ or $R$  for $N=2m, m \in 
\mathbb{Z^+}$ implies that corresponding to each of its $m$ eigenvalues
$\lambda$, there exists an eigenvalue $-\lambda$. Thus, the diagonal matrix
$M_d$ can always be arranged to have the form $M_d=diagonal (\lambda_1,
-\lambda_1,\dots, \lambda_m,-\lambda_m )$ by assuming a particular ordering
of eigenvalues. Under the transformation generated by the orthogonal matrix
$\hat{O}$, the canonical variables $X, P$ and the column matrix $F$ transform
as follows:
\bea
{\hat{X}} &=& \hat{O} X, \ {\hat{P}} = \hat{O} P, \ {\hat{\Pi}} = \hat{O} \Pi
={\hat{P}} + {\cal{A F}}, \ \nonumber \\
{\cal{F}} & \equiv & \hat{O} F, \  {\cal{A}} \equiv \hat{O}A\hat{O}^T.
\label{newcoor}
\eea
The Hamiltonian $H$ in Eq. (\ref{H}) may now be written as
defined in the background of an indefinite metric $M_d$:
\bea
H & = &\hat{\Pi}^T M_d \hat{\Pi} + V({\hat{x}}_1, {\hat{x}}_2, \dots, 
{\hat{x}}_N)\nonumber \\
& = & {\hat{P}}^T M_d {\hat{P}} + {\cal{F}^T} {\cal{A}}^TM_d {\hat{P}} +
{\hat{P}}^T M_d {\cal{A}}{\cal{F}}+
{\cal{F}}^T {\cal{A}}^TM_d {\cal{A}}{\cal{F}}
+ V(\hat{x}_1, \dots, \hat{x}_N).
\label{newhamil}
\eea
This form of the Hamiltonian as defined in the background of a pseudo-Euclidean
metric is used in the later part of the article when translationally and
rotationally symmetric systems are discussed.\\

\subsection{Representation of matrices}

\noindent A particular realization of the matrices $M$, $R$ and $D$
satisfying the condition (\ref{1con}) for an $N=2m$ dimensional
system may be obtained as follows:
\bea
M=I_m\otimes\sigma_x,\ \ \  \ A=\frac{-i\gamma}{2}I_m\otimes\sigma_y,\ \
D=\gamma \chi_m\otimes \sigma_z, \ \ [\chi_m]_{ij}=\frac{1}{2}\delta_{ij} Q_i(x_1, x_2,
\dots, x_N),
\label{repre}
\eea
\noindent where $\sigma_x, \sigma_y, \sigma_z$ are Pauli matrices and $I_m$ is
$m \times m$ identity matrix. The $m \times m$ diagonal matrix $\chi_m$ 
contains $m$ functions $Q_i$ which are dependent on specific choices of
$J$ and hence, on the functions $F_i$'s. In particular, the following equation may be
obtained by using Eqs. (\ref{repre}) and (\ref{1con}), 
\be
\chi_m \otimes I_2 = \frac{1}{2} J + \frac{1}{2} \left \{ I_m \otimes \sigma_y
\right \} J^T \left \{ I_m \otimes \sigma_y \right \}.
\label{chicond}
\ee
\noindent The choice of the matrix $D$ implies that the balancing of loss
and gain terms occur between the $(2i-1)^{th}$ and the $(2i)^{th}$
particles. It may be assumed at this point that space-dependent gain/loss terms 
for $(2i-1)^{th}$ and $(2i)^{th}$ particles solely depend on the co-ordinates
$x_{2i-1}$ and $x_{2i}$. Each particle may interact with rest of the particles
through the potential $V \equiv V(x_1, x_2, \dots, x_N)$. This scheme is
implemented by choosing,
\be
F_{2i-1} \equiv F_{2i-1}(x_{2i-1},x_{2i}),\ \ F_{2i} \equiv F_{2i}(x_{2i-1},x_{2i}) \ i=1, 2, \dots, m,
\ee
\noindent which implies that $J$ takes a block-diagonal form:
\be
J= \sum_{i=1}^m U_i^{(m)} \otimes V_i^{(2)}, 
\label{j1}
\ee
\noindent where $m$ number of $m \times m$ matrices $U_a^{(m)}$ and
$2 \times 2$ matrices $V_a^{(2)}$ are defined as,
\be
\left [ U_a^{(m)} \right ]_{ij} \equiv \delta_{ia} \delta_{ja}, \ \
V_a^{(2)} \equiv
\bp
{\frac{\partial F_{2a-1}}{\partial x_{2a-1}}}
& {\frac{\partial F_{2a-1}}{\partial x_{2a}}}\\
{\frac{\partial F_{2a}}{\partial x_{2a-1}}} &
{\frac{\partial F_{2a}}{\partial x_{2a}}}
\ep.
\label{j2}
\ee
\noindent Substituting Eqs. (\ref{j1}) and (\ref{j2}) in Eq. (\ref{chicond}),
$Q_a$ is determined as,
\be
Q_a(x_{2a-1},x_{2a}) = Trace (V_a^{(2)}),
\ee
which completely specifies the representation. For the case of constant balanced loss and gain coefficients,
$F_{2i-1}=x_{2i-1}$ and $F_{2i}=x_{2i}$ and the result of Ref. \cite{pkg-ds} are reproduced.

With this particular representation, the Hamiltonian $H$ takes the following
form,
\bea
H=\sum_{i=1}^m \left[2P_{2i-1}P_{2i}+\gamma \left(F_{2i-1}P_{2i-1}-F_{2i}P_{2i}\right)-\frac{\gamma^2}{2}F_{2i-1}F_{2i}\right]+V(x_1,x_2,\dots,x_N),
\label{Ham1}
\eea
where the conjugate momenta are given as:
\bea
P_{2i-1}=\frac{1}{2}(\dot{x}_{2i}+\gamma F_{2i}),\ \ \ \ \ P_{2i}=\frac{1}{2}(\dot{x}_{2i-1}-\gamma F_{2i-1}).
\eea
The equations of motion have the following form
\bea
\ddot{x}_{2i-1} - \gamma Q_i \dot{x}_{2i-1}+
2\frac{\partial V}{\partial x_{2i}}&=&0,\nonumber \\
\ddot{x}_{2i} + \gamma Q_i \dot{x}_{2i}+
2\frac{\partial V}{\partial x_{2i-1}}&=&0,
%\ \ \ Q= \frac{\partial F_{2i-1}}{\partial x_{2i-1}}+
%\frac{\partial F_{2i}}{\partial x_{2i}},
\label{equmo}
\eea
which, in general, constitute a set of $2m$ coupled Li$\acute{e}$nard-type
differential equations. The choice of a general quadratic form for $V$,
\be
V= X^T S X, \ S^T=S,
\ee
gives a chain of coupled linear oscillators with space-dependent balanced
loss and gain. Various physical situations may be taken into account by
choosing the symmetric matrix $S$ appropriately. An analytical solutions of this
system becomes nontrivial, since presence of space-dependent loss/gain
co-efficient makes the system nonlinear. A chain of non-linear oscillators
may also be constructed by appropriately choosing $V$. In general, finding
exact solutions of such systems are nontrivial. A few examples of exactly
solvable models with stable bound solutions will be discussed later in this
article.

\subsection{Unidirectional coupling between system and bath}

There is no coupling between the  dissipative and its auxiliary
system in the case of Bateman oscillators. The dynamics of the system can
be studied analytically both at the classical as well as quantum level. The
situation changes significantly if nonlinear terms are incorporated in the
system through the potential and analytical treatment seems nontrivial for
the corresponding classical as well quantum system. It is worth enquiring
at this juncture whether or not the tools and techniques associated with a
Hamiltonian formulation can be used to study the purely dissipative dynamics.
As a first step in this direction, it is required to choose the potential $V$
and $Q_i$ suitably such that the particles subjected to dissipative
dynamics are coupled among themselves only. However, the dynamics governing
the particles associated with auxiliary system may depend on dynamics of all
the particles. Thus, a kind of `unidirectional coupling' is required which may
be obtained for the following choices of $V$ and $Q_i$,
\bea
V=\sum_{i=1}^m x_{2i} V_i(x_1, x_3, \dots, x_{2m-1}), \ \
Q_i \equiv Q_i(x_{2i-1}), \ i=1, 2, \dots m,
\label{uni}
\eea
where $V_i$'s couple odd-numbered particles only. In this case
Eqs. (\ref{equmo}) reduce to following form:
\bea
&& \ddot{x}_{2i-1}-\gamma Q_i(x_{2i-1}) \dot{x}_{2i-1}+2V_i(x_1, x_3,\dots,
x_{2m-1})=0,\nonumber \\
&& \ddot{x}_{2i}+\gamma Q_i(x_{2i-1}) \ \dot{x}_{2i}+2\sum_{j=1}^m x_{2j}
\frac{\partial V_j}{\partial x_{2i-1}}=0.
\label{equmo1}
\eea
The odd-numbered particles interact among themselves, while the even-numbered
particles interact with all the particles for generic $V_j$. A Hamiltonian
formulation in its standard form is not possible involving either only
odd-numbered or even-numbered particles. However, the odd and even-numbered
particles together form a Hamiltonian  system.  One interesting observation
at this point is that the dynamics of the even-numbered particles are governed
by $m$ linear non-autonomous equations. The time-dependent co-efficients are
determined by solutions of the odd-numbered particles. For the specific
choice of $V_i \equiv V_i(\{x_{2i-1}\})$, the dynamics of odd-numbered particles
is governed by $m$ decoupled Li$\acute{e}$nard equations. Exact solutions
of Li$\acute{e}$nard equations for specific forms of $Q_i$ and $V_i$ are
known\cite{harko} which may be used to find the analytical solutions for the
even numbered particles.

The main advantage of systems with `unidirectional' coupling is that the
tools and techniques associated with Hamiltonian formulation like,
canonical perturbation theory, canonical quantization, KAM theory etc. may
be used to study the dynamics of purely dissipative systems. For example,
canonical perturbation theory may be used to study the dynamics of odd-numbered
particles for the choices of $Q_i(x_{2i-1})$ and $V_i(\{x_{2i-1}\})$ for which
an analytical solution is not possible. Such an investigation for $N=2$ has been
carried out for the case of Van der Pol oscillator\cite{sagar}. The generic
many-particle Hamiltonian $H$ with `unidirectional' coupling specified by
Eqs. (\ref{uni}) and (\ref{equmo1}) may be used to study the dynamics of
any dissipative system. For example, the choice of $V_i$ as,
\bea
&& V_i = \frac{\omega^2}{2} x_{2i-1} - \sum_{\substack {j=1\\(j \neq i)}}^m
\frac{g}{(x_{2i-1}-x_{2j-1})^3},\nonumber \\
&& V=\frac{\omega^2}{2} \sum_{i=1}^m x_{2i} x_{2i-1} -
g \sum_{\substack {i,j=1\\(j \neq i)}}^m
\frac{x_{2i}}{(x_{2i-1}-x_{2j-1})^3},
\eea
\noindent gives a dissipative rational Calogero model:
\bea
&& \ddot{x}_{2i-1}-\gamma Q_i(x_{2i-1}) \dot{x}_{2i-1}+
\omega^2 x_{2i-1} - 2 \sum_{\substack {j=1\\(j \neq i)}}^m
\frac{g}{(x_{2i-1}-x_{2j-1})^3}
=0,\nonumber \\
&& \ddot{x}_{2i}+\gamma Q_i(x_{2i-1}) \ \dot{x}_{2i}+
\omega^2 x_{2i} + 6 g \sum_{\substack {j=1\\(j \neq i)}}^m
\frac{x_{2i}-x_{2j}}{(x_{2i-1}-x_{2j-1})^4}=0.
\label{calo-dissi}
\eea
The equations of motion for the odd numbered particles in the limit $\gamma=0$
is identical with that of rational $A_{N+1}$-type Calogero model
\cite{calo,sut, ob,poly,pkg1} with $m$ particles. However,
the Hamiltonians for these two cases are not identical in the same limit, due
to a mismatch of total degrees of freedom and $\gamma$ independence of $V$. 
It may be noted that the quadratic terms in $V$ correspond to harmonic
confinement, while the terms with coefficient $g$ scales inverse-squarely as
in the case of rational Calogero model. However, the potential $V$ is not permutation symmetric.
Thus, the potential for the rational Calogero model and that of $H$
share some of the properties, although they are not identical. The integrability
and/or solvability of this system is not apparent, unlike the case of
Calogero-type Hamiltonian considered in Ref. \cite{pkg-ds}. An approximate
description is possible both at the classical as well quantum level by treating
$\gamma$ as a perturbation parameter and $H$ in Eq. (\ref{Ham1}) with
$\gamma=0$ as unperturbed Hamiltonian. The description of dissipative rational
Calogero model is left for future investigations. 
There are various many-particle integrable systems like Calogero-Sutherland models,
Toda lattice \cite{toda,toda1} etc. with interesting physical behaviours. These models appear in diverse branches of physics
from condensed matter systems to high energy physics \cite{quch, integ, pkg2,cfmp, tl,sata,tyama}.
A generalization of these models by including dissipation and investigating this new class of models is desirable. As a first
step towards this direction, a Hamiltonian formulation of these celebrated models with balanced loss and gain
can be obtained by employing the above formulation.

\subsection{Hamiltonian on a pseudo-Euclidean plane}

A Hamiltonian system with balanced loss and gain can always be reformulated
as a many-particle system in the background of a pseudo-Euclidean metric. Such
a construction is presented in this section.
For the representation (\ref{repre}), the orthogonal matrix
$\hat{O}$ that diagonalizes $M$ has the form
$
\hat{O}=\frac{1}{\sqrt{2}}
\left[I_m\otimes \left(\sigma_x+\sigma_z\right)\right]
$
and the matrices $M_d$ and $\cal{A}$ are respectively given by, 
\bea
M_d=I_m\otimes \sigma_z,\ \ {\cal{A}}=\frac{i\gamma}{2} \left(I_m\otimes \sigma_y\right).
\eea
Denoting the new canonical variables $\hat{X}$ and $\hat{P}$ by
$\hat{X}=(z_i^+,z_i^-)^T$ and $\hat{P}=(P_{z_i^+},P_{z_i^-})^T$ with
$i=1,2,\dots ,m$, the Hamiltonian of Eq. (\ref{newhamil}) may now be written as,
\bea
H&=&\sum_{i=1}^m\left[\left(P^2_{z_i^+}-P^2_{z_i^-}\right)+\gamma\left(F_i^+P_{z_i^-}+F_i^-P_{z_i^+}\right)-\frac{\gamma^2}{4}
\left\{(F_i^+)^2-(F_i^-)^2 \right\}\right]+V(\{z_i^+,z_i^-\}),\nonumber\\
&=&\sum_{i=1}^m\left[\left(P_{z_i^+}+\frac{\gamma}{2}F_i^-\right)^2-\left(P_{z_i^-}-\frac{\gamma}{2}F_i^+\right)^2\right]
+V(\{z_i^+,z_i^-\}),
\label{hz}
\eea
where the momenta conjugate to the variables $z_i^+,z_i^-$ are respectively given by

\bea
P_{z_i^+}&=&\frac{1}{2}\left(\dot{z}_i^+-\gamma F_i^- \right), \ \ \ \ P_{z_i^-}=-\frac{1}{2}\left(\dot{z}_i^--\gamma F_i^+ \right).
\eea
The relation between the old and new variables given by Eq. (\ref{newcoor})
may be expressed as
\bea
z_i^+&=&\frac{1}{\sqrt{2}}(x_{2i-1}+x_{2i}), \ \ \ \ \ z_i^-=\frac{1}{\sqrt{2}}(x_{2i-1}- x_{2i}).
\label{transzzz}\\
F_i^+&=&\frac{1}{\sqrt{2}}\left(F_{2i-1}+F_{2i}\right), \ \ \ \ \ F_i^-=\frac{1}{\sqrt{2}}\left(F_{2i-1}-F_{2i}\right).
\eea
It should be mentioned here that since $F_{2i-1}, F_{2i}$ are functions of $(x_{2i-1}, x_{2i})$, it is apparent that $F_i^+,F_i^-$
are functions of $(z_i^+,z_i^-)$ variables only, i.e., $F_i^+=F_i^+(z_i^+,z_i^-)$ and $F_i^-=F_i^-(z_i^+,z_i^-)$.
For the special case $F_i=x_i, \forall  \ i$, $F_i^+$ and $F_i^-$ respectively become $z_i^+$ and $z_i^-$ which
obviously  describe a system with constant balanced loss and gain coefficients.
The equations of motion corresponding to the Hamiltonian (\ref{hz}), take the following form:

\bea
\ddot{z}_i^+-\gamma Q_i(z^+_i,z^-_i)\dot{z}_i^-+2\frac{\partial V}{\partial z_i^+}&=&0,\nonumber \\
\ddot{z}_i^--\gamma Q_i(z^+_i,z^-_i)\dot{z}_i^+-2\frac{\partial V}{\partial z_i^-}&=&0,
\ \ Q_i=\left(\frac{\partial F^+_i}{\partial z_i^+}+\frac{\partial F^-_i}{\partial z_i^-}\right).
\label{eqz}
\eea
The equations (\ref{eqz}) may also be obtained from the Lagrangian $L$
corresponding to the Hamiltonian (\ref{hz}):
\bea
L=\sum_{i=i}^m\left[\frac{1}{4}\left\{(\dot{z}_i^+)^2-(\dot{z}_i^-)^2\right\}+\frac{\gamma}{2}\left(\dot{z}_i^-F_i^+-\dot{z}_i^+F_i^-\right)\right]-V(\{z_i^+,z_i^-\}).
\label{Lz}
\eea
The Hamiltonian in Eq. (\ref{hz}) may be interpreted as a system of $m$ 
particles on a pseudo-Euclidean plane with the metric $g_{ij}=(-1)^{i+1}
\delta_{ij}, i,j=1, 2$ and the $i$th particle being subjected to an external
inhomogeneous magnetic field $Q_i$. It may be noted that in the original
co-ordinate system defined by $x_i$'s, the space-dependent
gain/loss coefficient for the $(2i-1)$th and $(2i)$ th particles is also
$Q_i$. The gauge transformation
of the vector potential corresponding to external magnetic field produces
a Lagrangian differing from $(\ref{Lz})$ by a total time derivative term.
This point is elaborated further while quantizing the system and in the
Appendix-A. 

It should be mentioned here that a slight modification of the
Lagrangian of Eq. (\ref{Lz}) of the form,
\bea
L_t=\sum_{i=i}^m\left[\frac{1}{4}\left\{(\dot{z}_i^+)^2-(\dot{z}_i^-)^2\right\}+\frac{\gamma}{2}
\left(\dot{z}_i^-F_i^+-\dot{z}_i^+F_i^-\right)+\left(z_i^+h^+_i(t)+z_i^-h^-_i(t)\right)\right]-V(\{z_i^+,z_i^-\}),
\label{Lz1}
\eea
where $h^+_i(t)$,  $h^+_i(t)$ are arbitrary functions of time, will incorporate
a system with space dependent balanced loss/gain term, which is externally
driven. The Hamiltonian corresponding to the Lagrangian (\ref{Lz1}) has the
following form:
\bea
H_t=\sum_{i=1}^m\left[\left(P_{z_i^+}+\frac{\gamma}{2}F_i^-\right)^2-
\left(P_{z_i^-}-\frac{\gamma}{2}F_i^+\right)^2
-\left\{z_i^+h^+_i(t)+z_i^-h^-_i(t)\right\}\right]
+V(\{z_i^+,z_i^-\}).
\label{Ht}
\eea
A proper choice of the functions $h_i^+(t), h_i^-(t)$ can be made such that
only the particles associated with the chosen degree of freedom are externally driven. The explicit time dependence
of the Hamiltonian spoils the conservative nature of the system and will not
be considered further for discussions. The main emphasize of the present
article is to investigate the integrability and exact solvability of systems
with space-dependent balanced loss and gain terms. Such systems characterized
by translational or rotational symmetry are considered in the next two
sections.

\section{Translationally invariant system}

This section deals with the system described by the Hamiltonian (\ref{hz}) when the potential admits a translational
symmetry. It should be noted that under a constant and equal amount of shift of the coordinates $(x_{2i-1},x_{2i})$ of
the form $x_{2i-1}\rightarrow x_{2i-1}+\eta_i$ and $x_{2i}\rightarrow x_{2i}+\eta_i$, where $\eta_i$'s are $m$ independent
parameters, the action ${\cal A}=\int L dt $ remains invariant or differs at most by a total time derivative provided that 
the potential is only a function of $z_i^-$, i.e. $V= V(\{z_i^-\})$ and $Q_i= Q_i(z_i^-)$. 
%\bea
%\frac{\partial F_i^+}{\partial z_i^+}+\frac{\partial F_i^-}{\partial z_i^-}
%=Q_i(z_i^-).
%\label{con}
%\eea 
%It should be mentioned here that although the condition (\ref{con}) imposes a
%restriction on the possible form of $F_i^+$ and $F_i^-$, still there exist an infinite
%number of solutions for  $F_i^+$ and $F_i^-$ for a particular form of $Q_i(z_i^-)$.
For translationally symmetric system the first set of Eqs. (\ref{eqz}) can be solved
to give:

\bea
\dot{z}_i^+=\gamma f_i(z_i^-)+\Pi_i,\  \Pi_i \in \Re, \forall \ i,  \ \ \ \ \  f_i=\int Q_i(z_i^-)dz_i^-,
\label{dotz1}
\eea

\noindent where $\Pi_i$'s are $m$ integration constants to be determined by fixing the
initial conditions. Substituting the expression of $\dot{z}_i^+$ from Eq. (\ref{dotz1}) to the second set of
Eq. (\ref{eqz}), the following decoupled equation is obtained  for the variables $z_i^-$:

\bea
%\ddot{z^-}-\gamma^2F^- \frac{\partial F^-}{\partial z^-}\gamma \Pi \frac{\partial F^-}{\partial z^-}-2\frac{\partial V}{\partial z^-}=0,\\
\ddot{z}_i^-=\frac{dg_i}{dz_i^-},\ \ \ \ g_i=\frac{\gamma^2}{2}f_i^2+\gamma \Pi_i f_i +2 V(\{z_i^-\}),\ \ i=1,2,\dots,m.
\label{0eqz1}
\eea
Integrating Eq. (\ref{dotz1}) the following expression is obtained
for the variables $z^+_i$,

\bea
z_i^+=\gamma \int f_i dt +\Pi _it + C_i, \ \ C_i \in \Re, \forall \ i, 
\label{z1}
\eea
where $C_i$ are constants of integration. It is evident that for nonzero $\Pi_i$, 
the solutions of $z_i$ contain a linear dependence on time which introduces 
instability in the system. Therefore, in order to have stable solution for $z_i^+$,
$\Pi_i$ must be taken to be zero.
 It should be mentioned here that the 
$\Pi_i$'s appearing in Eq. (\ref{dotz1}) are $m$ integrals of motion:

\bea
\Pi_i=2P_{z^+_i}+\gamma \left(F^-_i-f_i\right),\ \ \ \{H,\Pi_i\}_{PB}=0\ \ \  \{\Pi_i,\Pi_j\}_{PB}=0,
\eea
where $\{,\}_{PB}$ denotes Poisson bracket. 
Therefore, the existence of $m+1$ integrals of motion $H,\Pi_i, \forall \ i$ in
involution, implies that the system described by the translationally invariant potential
where the space dependent balanced loss and gain coefficients are only functions of $z^-_i$
variables, is at least partially integrable. The existence of $m$ integrals of motion
are due to the invariance of the Hamiltonian under translations with $m$ independent
parameters $\eta_i$. For the potential of the form $V=V(\{z_i^+\})$, the action ${\cal A}=\int L dt $ remains
invariant or differs at most by a total time derivative under the translations $z_i^- \rightarrow z_i^-+\eta_i$ provided $Q_i= Q_i(z_i^+)$.
The corresponding $m$ conserved quantities
$
\Pi^-_i=-2P_{z_i^-}+\gamma\left(F_i^+-f_i^+\right), \   f^+_i=\int Q^+_i(z_i^+)dz_i^+,
$
and the Hamiltonian $H$ are in involution, implying that the system is partially integrable.
Further discussions in this article will be restricted to the case $V=V(\{z_i^-\})$. A few
choices of $V$ for which exactly solvable models can be constructed are presented below.

\subsection{Solution for two dimensional system}

As a simple example, the two dimensional case with $m=1$ and $N=2$ is presented in this subsection.
The two dimensional system is completely integrable with $H$ and $\Pi_1$ being two
integrals of motion in involution.
 The potential $V$ and the function $f_1$ are chosen
to have the form
\bea
V&=&-\frac{\omega_0^2}{4}(z_1^-)^2-\frac{\alpha_0}{6}(z_1^-)^3-\frac{\beta_0}{8}(z_1^-)^4, \  \  \omega_0,\alpha_0,
\beta_0 \in \Re,\\
f_1&=&a z_1^-+\frac{b}{\sqrt{2}}(z_1^-)^2,\ \  a,b \in \Re.
 \eea
In this case Eq. (\ref{0eqz1}) becomes:

\bea
\ddot{z}_1^-&+&\omega^2 z_1^-+\alpha (z_1^-)^2+\beta (z_1^-)^3=\gamma \Pi_1 a, \nonumber\\
\omega^2&=&\omega^2_0-\gamma(\sqrt{2}\Pi_1 b+\gamma a^2), 
\ \ \  \alpha=(\alpha_0-\frac{3}{\sqrt{2}}ab\gamma^2), \ \ \  \beta=\beta_0-\gamma^2b^2.
\label{transz2}
\eea
There exists various choices of the parameters for which exact solutions can be constructed.
As an example the following  case with $\Pi_1=0, \alpha=0$ may be considered.
In this case $\alpha_0=\frac{3}{\sqrt{2}}ab\gamma^2$ and Eq. (\ref{transz2}) becomes
 
 \bea
\ddot{z}_1^-&+&\omega^2 z_1^-+\beta (z_1^-)^3=0,  \nonumber\\
\omega^2&=&\omega^2_0-\gamma^2 a^2, 
\ \ \  \beta=\beta_0-\gamma^2b^2,
\label{transz23}
\eea
the solutions of which are given by the quartic oscillator and are
discussed in Refs.\cite{pkg-ds,ml}.\\
 
\noindent  i) $\omega^2 > 0, \beta >0$: In this case the parameter $\gamma$ is restricted
to lie in the range, $-\frac{\omega_0}{a} < \gamma < \frac{\omega_0}{a}$ and $\beta_0>\gamma^2 b^2,\beta_0>0$. 
The solution for $z_1^-$ is given by 

\bea
z_1^-(t)= A \ cn (\Omega t,k),\ \ 
\Omega=\sqrt{\omega^2 + \beta A^2}, \
k^2=\frac{\beta A^2}{2 \Omega^2}.
\label{solex1}
\eea
For non-singular stable solutions, the range of $k$ is $0< k < 1$.
It should be noted that Eq. (\ref{transz2}) is a second order ordinary differential equation
and therefore, contains two integration constants. In this case $A$ is one of the integration constants
and the other integration constant appearing as a phase of Jacobi elliptical function is taken to be zero.
This can always be done by fixing the position of the particle at $t=0$.
The solution for $z_1^+$ is obtained from Eq. (\ref{z1}) and has the form
\bea
z_1^+&=&
\frac{bA^2\gamma}{\Omega\sqrt{2}}
\left[\Omega t-\frac{\Omega t}{k}+\frac{E [am[\Omega t,k],k]
 (-1+\frac{1}{k}+cn^2[\Omega t, k])}{dn[\Omega t,k]\sqrt{1-ksn^2[\Omega t,k]}}\right]\nonumber\\
 &+&
\frac{a A\gamma}{\Omega} \frac{\cos^{-1} \{ dn(\Omega t,k) \}
sn(\Omega t,k)}{\sqrt{1-dn^2(\Omega t,k)}}+C_1,
\label{solex0}
\eea
where $E$ denotes elliptic integral of second kind and $C_1$ is the constant of integration.

\noindent (ii) $\omega^2 > 0, \beta < 0$: The parameter $\gamma$ is restricted
to lie in the range,  $-\frac{\omega_0}{a} < \gamma < \frac{\omega_0}{a}$ and
$\beta_0<\gamma^2 b^2$ for $\beta_0>0$.
The solution for $z_1^-$ is given by

\bea
 z_1^-(t) =  A \ sn(\Omega t,k),\ \
\Omega=(\omega^2 -\frac{{\mid \beta \mid} A^2}{2})^{\frac{1}{2}}, \
k^2 = \frac{{\mid \beta \mid} A^2}{2 \Omega^2 }, \
0 \leq A \leq \sqrt{\frac{\omega^2}{{\mid \beta \mid}}}.
\label{solex2}
\eea
For non-singular stable solutions, the range of $k$ is $0< k < 1$.
The solution for $z_1^+$ has the form

\bea
z_1^+&=&
\frac{bA^2\gamma}{\sqrt{2}\Omega k}\left[\Omega t-\frac{E[am[\Omega t,k],k]
 \sqrt{(1-ksn^2[\Omega t, k])}}{dn[\Omega t,k]}\right]\nonumber\\
&+& \frac{a A\gamma}{\sqrt{k} \Omega} log \left [ dn(\Omega t,k)
-\sqrt{k} cn(\Omega t,k) \right ]+C_1,
\eea
where $C_1$ is the constant of integration.

\noindent (iii)  $\omega^2 < 0, \beta> 0$: In this case the angular
frequency is restricted to lie in the range, $-a\gamma < \omega_0 < a\gamma$
 and $\beta_0>\gamma^2 b^2,\beta_0>0$. Depending upon the range of the amplitude $A$,
 two solutions are obtained. One of which is stable and another is unstable. The unstable solution for $z^-_1$ 
 has the form

 \bea
 z_1^-(t) =  A dn(\Omega t,k),\ \ 
 \Omega=(\frac{{\beta } A^2}{2}), \
k^2 = \frac{\beta A^2-{\mid \omega^2 \mid}}{2 \Omega^2 }, \
\sqrt{\frac{{\mid \omega^2 \mid}}{\beta}} \leq A \leq 
\sqrt{\frac{2 {\mid \omega^2 \mid}}{\beta}},
\label{solex3}
\eea
\noindent where $z_1^+(t)$ is unbounded for the range $0 < k < 1$. 
The solution for $z^+_1$ reads

 \bea
z_1^+=\frac{bA^2\gamma}{\Omega\sqrt{2}}\left[\frac{E[am[\Omega t,k],k]
 dn[\Omega t, k])}{\sqrt{1-ksn^2[\Omega t,k]}}\right]
 + \frac{a A \gamma}{\Omega} am(t,k)+C_1,
 \eea
where $C_1$ is the constant of integration. 
The stable solution for $z^-_1$ is obtained for $\sqrt{\frac{2 {\mid \omega^2 \mid}}{\beta}}
\leq A < \infty $ which is similar to Eq. (\ref{solex1}) except for the
expressions for $\Omega$ and $k$ and has the form
\bea
 z_1^-(t)= A \ cn (\Omega t,k),\ \
 \Omega= \sqrt{-{\mid \omega^2 \mid} + \beta A^2}, \
k^2=\frac{\beta A^2}{2 \Omega^2}. 
\label{solex}
\eea
 In this case the solution for $z_1^+$ is
 given by Eq.(\ref{solex0}) and the expression for $\Omega$ and $k$ is given by Eq.(\ref{solex}).
 The solutions obtained for case-I are not stable. Some observations on the obtained solutions 
 are as follows:
 
 \noindent i) For the choice $\alpha_0=b=0$, $f_1=a z^-_1$ which is the case of the constant gain-loss 
 coefficients and the solutions for $z^-_1$ are that of a quartic oscillators. In this case all the solutions as discussed for the
 case-I reduce to the form as obtained in Ref. \cite{pkg-ds}. The solutions are stable in this case.
 
 \noindent ii) For $\alpha_0=a=0$, $f_1$ becomes $\frac{b}{\sqrt{2}}(z^-_1)^2$ which corresponds to the 
 case of linear loss-gain coefficients. The solutions for $z^-_1$ is again given by the solutions of a
 quartic oscillators, with only a change in the parameters range. In this case the solutions 
 for $z^-_1$ and $z^+_1$ can be obtained by putting $a=0$ in all solutions as obtained above 
 and taking the range of the parameters appropriately. However, the solutions obtained in this case are
 not stable.
 
 \noindent iii) It should be noted that for constant balanced loss and gain coefficients, 
 translationally symmetric systems admit stable solutions. However, the introduction of space
 dependent balanced loss and gain coefficient makes the system unstable for the same
 form of the interacting potential.\\

\section{Rotationally symmetric system:}

The system described by the Hamiltonian (\ref{hz}), with $J$ given by (\ref{j1}),
 may be considered as $m$ copies of a two dimensional system interacting with each other via 
 the potential $V$.
%potential $V$ admits a rotational symmetry in a pseudo Euclidean space.
For the choice of the functions $F_i^+=z_i^+g(r_i)$ and $F_i^-=z_i^-g(r_i)$ where $r_i^2=(z_i^+)^2-(z_i^+)^2$, 
a set of $m$ constants of motion can be constructed for a class of potential $V=V(\{r_i\})$:

\bea
L_i=\left(z_i^-\dot{z}_i^+-z_i^+\dot{z}_i^-\right)+\gamma r_i^2g(r_i),\ \ i=1,\dots,m.
\eea
The $m$ conserved quantities $L_i, i=1,\dots,m$ are due to the  rotational symmetry
under rotation in a pseudo Euclidean space that each copy of $m$ two dimensional system possesses
when  the potential is a function of $\{r_i\}$ only, i.e, $V=V(\{r_i\})$. At this stage, a convenient choice
of the coordinates of the form

\bea
z_i^+=r_i \cosh{\theta_i},\ \ \ \  z_i^-=r_i \sinh{\theta_i},
\label{rtcoor}
\eea
cast the Hamiltonian of Eq. (\ref{hz}) in the following form

\bea
H=\sum_{i=1}^m\left[P_{r_i}^2-\frac{1}{r_i^2}\left(P_{\theta_i}-\frac{1}{2}\gamma r_i^2 g \right)^2\right]+V,
\label{hyh}
\eea
where $P_{r_i}$ and $P_{\theta_i}$ are respectively the momenta conjugate to $r_i$ and $\theta_i$ coordinates:

\bea
P_{r_i}=\frac{\dot{r}_i}{2},\ \ \ \ 
P_{\theta_i}=-\frac{1}{2}\left(r_i^2\dot{\theta}_i-\gamma r_i^2 g\right)=\frac{L_i}{2}.
\eea
It should be mentioned that the Hamiltonian in Eq. (\ref{hyh}) becomes independent of $\theta_i$ for 
$V=V(\{r_i\})$ and the momentum $P_{\theta_i}$ conjugate to $\theta_i$ becomes a constant of motion.
Further, the result $\{H, P_{\theta_i}\}_{PB}=0$ implies the existence of $m$ integrals of motion
in convolution which indicates that the system is at least partially integrable.
The equations of motion corresponding to the $r_i$ and $\theta_i$ variables are respectively:

\bea
&\ddot{r}_i&+\frac{4}{r_i^3}P^2_{\theta_i}-\gamma^2 r_i g^2 +\gamma\left(2 P_{\theta_i}-\gamma r_i^2 g\right)\frac{\partial g}{\partial r_i}
+2\frac{\partial V}{\partial r_i}=0,
\label{eqr}\\
&\dot{\theta}_i&=\frac{1}{r_i^2}\left(\gamma r_i^2 g-2P_{\theta_i}\right).
\label{eqtheta}
\eea

\subsection{Solution for two dimensional system}

In case of two dimensional system $m=1$ and $N=2$ and the system is completely
integrable since there exits two integrals of motion $H, P_{\theta}$ in involution. The following
cases are considered for a two dimensional system.\\

\noindent Case I: $P_{\theta}=0$, $g=$constant$=c$\\

\noindent This gives the case of constant balanced loss and gain coefficients \cite{pkg-ds}. 
In this case Eqs. (\ref{eqr}) and (\ref{eqtheta}) takes the following form:

 \bea
&\ddot{r}&-\gamma^2 r c^2 
+2\frac{\partial V}{\partial r}=0,\ \ \ \ \dot{\theta}=\gamma c.
\label{r}
\eea
It should be noted that in this case Eqs (\ref{eqr}) and (\ref{eqtheta}) are decoupled and the solution of 
$\theta$ is given as $\theta= c\gamma t +A$ with $A$ being a constant of
integration. The solutions for $z^+_1$ and $z^-_1$ may directly be written as

\bea
z^+_1=r(t)\cosh{( c\gamma t +A)},\ \ \ \ z^-_1=r(t)\sinh{( c\gamma t +A)}.
\eea 
The solutions in terms of $x_1$ and $x_2$ may be obtained from Eq.(\ref{transzzz}):

\bea
x_1=\frac{r(t)}{\sqrt{2}}\exp{\left[ c\gamma t +A\right]},\ \ \ \ 
x_2=\frac{r(t)}{\sqrt{2}}\exp{\left[-( c\gamma t +A\right)]}.
\label{x12}
\eea
Depending upon the form of $V(r)$, the solution for $r$ is obtained by solving Eq. (\ref{r}).
For example, in case $V=\frac{1}{4} \omega_0^2 r^2+\frac{\beta}{8}r^4$, the solutions
for $r$ is that of a quartic oscillator as has been discussed in section-3 and exact non-singular
solutions for $r$ can be found. It should be noted that the solutions for $x_1$ is always growing and
solutions for $x_2$ is always decaying as in the case of harmonic oscillator with balanced loss and gain
and without any coupling. Eq. (\ref{x12}) suggests that the introduction of any type of coupling for 
constant gain-loss coefficient in case of rotationally symmetric system in a pseudo Euclidean
 space with the constant of motion $P_{\theta}=0$, is unable to give any stable solutions.\\

\noindent Case II: $P_{\theta}=0$ \\

\noindent In this case Eqs. (\ref{eqr}) and (\ref{eqtheta}) takes the following form:

\bea
&\ddot{r}&-\gamma^2 r g \frac{d(rg)}{dr}+2\frac{\partial V}{\partial r}=0,\\
&\dot{\theta}&=\gamma g.
\eea

 \noindent Choice (i):  $g=cr, V=\frac{1}{4}\omega_0^2 r^2+\frac{1}{8}\alpha_0 r^4$.
 This choice gives the following equations for $r$ and $\theta$: 
 
 \bea
 &\ddot{r}&+\omega_0^2 r+\alpha r^3=0, \ \ \ \ \ \alpha=\alpha_0-2\gamma^2c^2,\\
 \label{qosci}
 &\dot{\theta}&=c\gamma r.
 \label{qoscila}
 \eea
 The solution of Eq. (\ref{qosci}) is given by the solution of a quartic oscillator and has been discussed in section-$3$.
 The solution for $\theta$ is obtained by integrating Eq.(\ref{qoscila}).
 For $\alpha_0>2\gamma^2c^2, \alpha_0>0$, the solutions for $r$ is given 
 by Eq. (\ref{solex1}). In terms of  $z_1^+$ and $z_1^-$ the solutions are
 
 \bea
 z_1^+&=&A cn[\Omega t,k]\cosh{\theta}, \ \ \ \ \ z_1^-= A cn[\Omega t,k]\sinh{\theta},\\
 \theta&=& \frac{c A\gamma }{\Omega} \frac{\cos^{-1} \{ dn(\Omega t,k) \}
sn(\Omega t,k)}{\sqrt{1-dn^2(\Omega t,k)}}+B, \ \ \ \ \
 \Omega=\sqrt{\omega_0^2 + \alpha A^2}, \
k^2=\frac{\alpha A^2}{2 \Omega^2},
 \eea
 where $B$ is a constant of integration.  In terms of $x_1$ and $x_2$ the solutions are
 \bea
 x_1=\frac{A cn[\Omega t,k]}{\sqrt{2}}\exp{(\theta)},\ \ \ \ 
x_2=\frac{A cn[\Omega t,k]}{\sqrt{2}}\exp{(-\theta)}.
\label{x1x2}
 \eea
 The solutions of $x_1$ and $x_2$ as given by Eq. (\ref{x1x2}) are non-singular stable and periodic. 
 For $\alpha_0<\gamma^2 c^2$, $\alpha_0>0$ or
$\alpha_0>\gamma^2 b^2$ for $\alpha_0<0$,
the solutions for $r$ is given by Eq. (\ref{solex2}).
 In terms of $z_1^+$ and $z_1^-$ the solutions are
 
 \bea
z_1^+&=&A \ sn(\Omega t,k)\cosh{\theta},\  z_1^-=A \ sn(\Omega t,k)\sinh{\theta},\nonumber\\ 
\theta&=& \frac{c A \gamma }{\sqrt{k} \Omega} log \left [ dn(\Omega t,k)
-\sqrt{k} cn(\Omega t,k) \right ]+B,\nonumber\\ 
 \Omega&=&(\omega_0^2 -\frac{{\mid \alpha \mid} A^2}{2})^{\frac{1}{2}}, \
k^2 = \frac{{\mid \alpha \mid} A^2}{2 \Omega^2 }, \
0 \leq A \leq \sqrt{\frac{\omega_0^2}{{\mid \alpha \mid}}},
 \eea
 where $B$ is a constant of integration.
  In terms of $x_1$ and $x_2$ the solutions are
 \bea
 x_1=\frac{A sn[\Omega t,k]}{\sqrt{2}}\exp{(\theta)},\ \ \ \ 
x_2=\frac{A sn[\Omega t,k]}{\sqrt{2}}\exp{(-\theta)}.
\label{x1x21}
 \eea
 The solutions of $x_1$ and $x_2$ as given by Eq. (\ref{x1x21}) are non-singular stable and periodic. 
 
\noindent  It is interesting to note that for constant balanced loss and gain coefficients, it is impossible to achieve
 stable solutions for any type of coupling in case of rotationally symmetric system in a pseudo Euclidean
 space when the constant of motion $P_{\theta}$ is taken to be zero. However, the introduction of space
 dependent balanced loss and gain coefficients makes it possible to achieve non-singular stable and periodic
 solutions in this case.

\section{Quantization of the classical Hamiltonian}
 
This section deals with the quantization of the classical Hamiltonian
$H$ in Eq. (\ref{hz}). The classical variables $P_{z_i^{\pm}}, z_i^{\pm}$
are treated as operators satisfying the standard commutation relations:
 \be
 \left [ z_j^+, P_{z_j^+}\right ] = i, \left [ z_j^-, P_{z_j^-} \right ] = i.
\label{commutator}
\ee
\noindent All other commutators involving $P_{z_j^{\pm}}$ and 
$z_j^{\pm}$ are taken to be zero. 
It may be noted that the canonical quantization method has been employed to quantize the system.
The classical Poission bracket
relations among the coordinates and the corresponding conjugate momenta are promoted to 
quantum commutators multiplied by the factor $\frac{1}{i\hbar}$ with the convention 
$\hbar=1$. Within the canonical quantization scheme,
it is also possible to quantize the same system by using guiding centre coordinates, 
since the gain/loss coefficient can be interpreted as analogous magnetic field. 
For such cases, the position operators become noncommutative. However, 
any such possibility is not considered in the present article.
 
A set of generalized momentum operators
$\Pi_{z_i^{\pm}}$ are introduced as follows:
\be
{\Pi}_{z_i^{\pm}}:= P_{z_i^{\pm}} - A_i^{\pm} =
-i \partial_{z_i^{\pm}} \pm \frac{\gamma}{2} F_i^{\mp}, \\
A_i^{\pm}:= \mp \frac{\gamma}{2} F_i^{\mp},
\ee
\noindent where the coordinate-space representation of the operators
$P_{z_j^{\pm}}$ is used, i.e. $P_{z_j^{\pm}}:=-i \partial_{z_j^{\pm}}$.
It may be recalled that $F_i^{\pm} \equiv F_i^{\pm}(z_i^-,z_i^+)$, which imply
the following commutation relations among the operators $\Pi_{z_i^{\pm}}$:
\be
\left [ \Pi_{z_i^{\pm}}, \Pi_{z_j^{\pm}} \right ] =0, \ \
\left [ \Pi_{z_i^{-}}, \Pi_{z_j^{+}} \right ] = -\delta_{ij}
\frac{i \gamma}{2} Q_i(z_i^-,z_i^+).
\label{mag-com}
\ee
\noindent Note that the appearance of space-dependent loss/gain coefficients
$Q_i(z_i^-,z_i^+)$ in the second set of commutation relations in
Eq. (\ref{mag-com}). The operators $A_i^{\pm}$ may be identified as two
dimensional vector potentials producing inhomogeneous magnetic fields
$Q_i(z_i^-,z_i^+)$ perpendicular to the `$z_i^--z_i^+$' planes. The case
of constant loss/gain co-efficient corresponds to uniform magnetic field.
For space-dependent loss/gain coefficients, a change in the direction of
magnetic field as a function of the co-ordinates corresponds to a change
in gain/loss experienced by the particle.

The quantum Hamiltonian $\hat{H}$ corresponding to $H$ in Eq. (\ref{hz})
has the following expression:
\bea
\hat{H}= \sum_{i=1}^m \left [ \left ( \Pi_{z_i^{+}} \right )^2
-\left ( \Pi_{z_i^{-}} \right )^2 \right ] + V(\{z_i^-,z_i^+\})
% &=& \sum_{j=1}^m  [\left(\partial^2_{z_j^-}-
%\partial^2_{z_j^+}\right)-i \frac{\gamma}{2} \left(F_{j}^{+}\partial_{z_j^-}+
%F_{j}^{-}\partial_{z_j^+}\right)- \frac{\gamma^2}{4}
%\left( ({F_j^+})^2-({F_j^-})^2 \right)\\
%&-&\frac{i\gamma}{2}\left(\frac{\partial F^+_i}{\partial z^-_i}+
%\frac{\partial F^-_i}{\partial z^+_i}\right) ]
%+V(z_j^-,z_j^+),
\label{quantumH}
\eea
\noindent where a symmetrization of the terms $F_i^+ P_{z_i^-} +
F_i^-P_{z_i^+}$ have been used. The  Hamiltonian can be interpreted as
a many-particle system defined in the background of a pseudo-Euclidean metric
with particles interacting with each other through the potential $V$ and
subjected to inhomogeneous magnetic field. There are provisions for writing
the quantum Hamiltonian $\hat{H}$ in different gauges, which at the classical
level corresponds to adding/subtracting total time-derivative terms to the
Lagrangian $L$. The following two unitary operators  are defined,
\be
S_1:= exp\left [ \frac{i \gamma}{2} \sum_{j=1}^m \int        
F_j^+(z_j^-,z_j^+) dz_i^- \right ],\ \ 
S_2:= exp\left [ \frac{i \gamma}{2} \sum_{j=1}^m \int 
F_j^-(z_j^-,z_j^+) dz_i^+ \right ], \ \
\ee
\noindent in order to elucidate the point.  The Hamiltonian
$\hat{H}$ may be transformed to unitary equivalent Hamiltonian
$\hat{H}_1:=S_1^{-1} H S_1$ and $\hat{H}_2:=S_2 H S_2^{-1}$ by using $S_1$
and $S_2$, respectively. In particular,
\bea
&& \hat{H}_1 = \sum_{i=1}^m \left [ 
\left ( P_{z_i^+} + \frac{\gamma}{2} 
\int Q_i(z_i^+,z_i^-) dz_i^-
 \right )^2  - P_{z_i^-}^2 \right ] + V(\{z_i^{-},z_i^+\}),\nonumber \\
&& \hat{H}_2 = \sum_{i=1}^m \left [ P_{z_i^+}^2 -
\left (P_{z_i^-} -\frac{\gamma }{2}  \int Q_i(z_i^+,z_i^-) dz_i^+\right )^2  \right ] + V(\{z_i^{-}, z_i^+\}).
\eea
\noindent Any one of the Hamiltonian $\hat{H}, \hat{H}_1, \hat{H}_2$
may be used depending on convenience and/or physical situations. In particular, the 
forms of $\hat{H}_1$ and $\hat{H}_2$ are suitable for box normalization \cite{ds-pkg1},which is required for
quantizing a translational invariant system.

\subsection{Translationally invariant system}

 %Some comments on this Hamiltonian are as follows:\\
 
 \noindent  The momentum operators $P_{z^+_i}$,
%\bea
% \hat{\Pi}_i=2P_{z^+_i}+\gamma \left(F^-_i-f_i\right),\ \ \ \ f_i=\int Q(z_i^-)dz_i^-, \ \ \ \ \  i=1,2,\dots,m,
 %\eea
commute with the Hamiltonian $\hat{H_1}$, provided $Q_i=Q_i(z_i^-)$, $V_i=V_i(\{z_i^-\})$.
In particular, the following commutation relations hold:
 \bea
 [H_1, P_{z^+_i}]=0,\ \ \ \ \ \  [P_{z^+_i},P_{z^+_j}]=0, \ \ \ \ \ i=1,2,\dots, m.
 \label{1com}
 \eea
 The existence of $m+1$ integrals of motion in involution imply the partial integrability
 of the system. However, the two-particle system is completely integrable.
The time-independent Schrodinger equation $H_1\psi=E\psi$ with 

\bea
\psi =\exp \left[ \sum_{j=1}^m i k_j z_j^+\right]\phi(\{z_i^-\}),
\eea 
takes the following form 
\bea
\sum_{i=1}^m\left[\partial^2_{z_i^-}+\left(k_i+\frac{\gamma}{2} f_i(z_i^-)\right)^2\right]\phi+V(\{z_i^-\})\phi=E \phi,
\label{sep}
\eea
where $k_i$'s are the eigenvalues of the operators $P_{z^+_i}$. Even for linear
space dependence of the gain-loss coefficients the functions $f_i$'s become
quadratic and the solution of Eq. (\ref{sep}) becomes nontrivial. The variable
$z_1$ will be used instead of $z_1^-$ in rest of this section for notational
convenience. For $N=2$
and $m=1$, some solutions corresponding to quasi-exactly solvable models are
presented with the following choices of the function $f_1$ and the potential
$V$,

\bea
V&=&-\alpha^2{z_1}^6 -\beta^2 {z_1}^2- 2\tilde{a} \tilde{b}{z_1}^4-2ab {z_1}^4+\tilde{a}\left(4n+2p+3\right)
{z_1}^2+\tilde{b}(1+2p),\nonumber\\
f_1&=&\frac{2}{\gamma}\left(a{z_1}^3+bz_1\right),\ \tilde{a}^2=\left(\alpha^2-a^2\right),\ \tilde{b}^2=\left(\beta^2-b^2\right).
\label{potf}
\eea
In this case Eq. (\ref{sep}) takes the following form:

\bea
-\partial^2_{z_1}\phi+V'\phi=-E \phi,
\label{sep1}
\eea
where $k_1$ is taken to be zero and $V'$ is given by:

\bea
V'(a,b)&=&\tilde{a}^2z_1^6+2\tilde{a}\tilde{b}z_1^4+\left\{\tilde{b}^2-\tilde{a}(4n+2p+3)\right\}z_1^2
-\tilde{b}(1+2p),
\eea
where the parameter $p$ takes the values $p=0,1$. 
The potential $V'$ is quasi-exactly-solvable when
$n$ is a non-negative integer and $\tilde{a}$ is non-negative number \cite{tur}. Since the potential $V'$ is even $V'(-z_1)=V'(z_1)$, the eigenfunctions 
can always be taken to have either even or odd parity. For even eigenfunctions $p$ is zero and for 
odd eigenfunctions $p$ is one. The Eq. (\ref{sep1}) for the potential $V'$ has the form of a sextic oscillator as discussed in Ref. \cite{tur}
with an exception that $E$ is replaced by $-E$. This change in sign manifests subtle issues of having well defined energy spectra and 
normalizable wavefunctions depending on the nature of the potentials \cite{ds-pkg1}. In order to address this issues for the
potential $V'$, the eigenfunctions and some of the energy eigenstates are presented below.

\noindent The solutions of Eq. (\ref{sep1}) may directly be written as \cite{tur},

\bea
\phi_{n}=z_1^pP_n(z_1^2)\exp{\left[-\frac{\tilde{b}}{2}z_1^2-\frac{\tilde{a}}{4}z_1^4\right]}, 
\label{phi}%\ i=0,1,\dots, n,
\eea
where $P_n$ is a polynomial of degree $n$ which is an element of $(n+1)$ dimensional representation of the $sl(2)$-
algebra. It should be mentioned that the models considered in Ref. \cite{ds-pkg1} also contain a negative sign in the 
right hand side of the time independent Schrodinger equation. In this case, in order that the systems possess normalizable wave functions
as well as an energy spectra which is bounded from below, proper Stoke wedges is needed to be defined where the wave functions
are normalizable. For example, in case $\tilde{a}=0$ the eigenfunctions and the 
energy spectra of the system described by Eq. (\ref{sep1}) is given by the eigenfunctions and the 
energy spectra of that of a harmonic oscillator but in this case in order to have an energy spectra which is bounded from below,
$\tilde{b}$ must be $\tilde{b}<0$ and the wavefunctions are not normalizable along the real $z_1$-axis. However, the wave functions are normalizable 
in the complex $z_1$-plane within the Stoke wedges of opening angle $\frac{\pi}{2}$ and centred about the positive and negative
imaginary axis \cite{aturcmb}. In the present case the wave functions (\ref{phi}) are normalizable 
along the real $z_1$-line due to the presence of the $-z_1^4$ term in the exponential with a coefficient $\tilde{a}>0$ and 
the normalization of the wavefunctions do not depend on the sign of the parameter $\tilde{b}$. 
 However, the wavefunctions (\ref{phi}) are also normalizable in the complex $z_1$-plane in Stoke wedges of opening angle $\frac{\pi}{4}$ 
and centred about the positive and negative real axis and in Stoke wedges of opening angle $\frac{\pi}{4}$ 
and centred about the positive and negative imaginary axis.
The normalization of the wavefunctions within the Stoke wedges of opening angle $\frac{\pi}{4}$ 
and centred about the positive and negative imaginary axis
is preferable since it produces the desire result in the case $\tilde{a}=0$ and $\tilde{b}<0$. 
If one substitutes $\phi_n$ from (\ref{phi}) to Eq. (\ref{sep1}), then
the following equation is obtained in the variable $y=z_1^2$,

\bea
-4y\frac{d^2P_n}{dy^2}+2\left(2\tilde{a}y^2+2\tilde{b}y-1-2p\right)\frac{dP_n}{dy}-4\tilde{a}nyP_n=-EP_n,
\label{po}
\eea
with $P_n=\sum_{j=0}^{n}c_jy^j$, the Eq. (\ref{po}) gives a system of $(n+1)$ linear homogeneous equations, the solution
of which gives the coefficients $c_j$. For non-trivial solutions of $c_j$'s, the determinant of the coefficients
must vanish. This determinant is a polynomial of degree $n+1$ in the variable $E$, the solutions of which determines the 
energy $E$ of the system.\\

\noindent For $n=0$:

\bea
V'(a,b)&=&\tilde{a}^2z_1^6+2\tilde{a}\tilde{b}z_1^4+\left\{\tilde{b}^2-\tilde{a}(2p+3)\right\}z_1^2-\tilde{b}(1+2p),\nonumber\\
E&=&0,\ \ \phi_{0}=z_1^p\exp{\left[-\frac{\tilde{b}}{2}z_1^2-\frac{\tilde{a}}{4}z_1^4\right]}.
\eea

\noindent For $n=1$:

\bea
V'(a,b)&=&\tilde{a}^2z_1^6+2\tilde{a}\tilde{b}z_1^4+\left\{\tilde{b}^2-\tilde{a}(2p+7)\right\}z_1^2-\tilde{b}(1+2p),\nonumber\\
E^{\pm}&=&-2\tilde{b}\pm 2 \left(\tilde{b}^2+2(1+2p)\tilde{a}\right)^{\frac{1}{2}},\nonumber\\
\phi_{1}&=&z_1^pP^{\pm}_1\exp{\left[-\frac{\tilde{b}}{2}z_1^2-\frac{\tilde{a}}{4}z_1^4\right]},\ \ \ \
P^{\pm}_1=2\tilde{a}z_1^2+\tilde{b}\pm\left(\tilde{b}^2+2(1+2p)\tilde{a}\right)^{\frac{1}{2}}.
\eea
It should be noted that in the limit $\tilde{a}=0$ and $\tilde{b}<0$, the results of Ref. \cite{ds-pkg1} are
reproduced when the normalization is carried out in the above mentioned Stoke wedges. The eigenvalues and
eigenfunctions obtained  in this case are different to that obtained in Ref. \cite{tur}. This is due to the negative sign
in the right hand side of Eq. (\ref{sep1}). It should be noted that the energy spectrum bounded from below  
and the corresponding normalized wave function are obtained only in the range $\tilde{a}>0$, $\tilde{b}\in \Re$.  
For $\tilde{a}<0$ the wave function (\ref{phi}) is not normalizable along the real $z_1$ line as well in the
Stoke wedges as discussed above. It should be mentioned here that for some particular models as discussed in 
Ref. \cite{ben,ds-pkg}, the quantum bound states occur at the same range of the parameters for which the classical
solutions are stable. However, no such result can be presented for the model under investigation, since the exact
classical solutions for the present model are not known and a linear stability analysis is inconclusive. Further
investigations by using nonlinear stability analysis may be required in order to get a conclusive result in this
regard.

\begin{figure}[h!]
\begin{center}
\fbox{\includegraphics[width=6cm,height=5 cm, ]{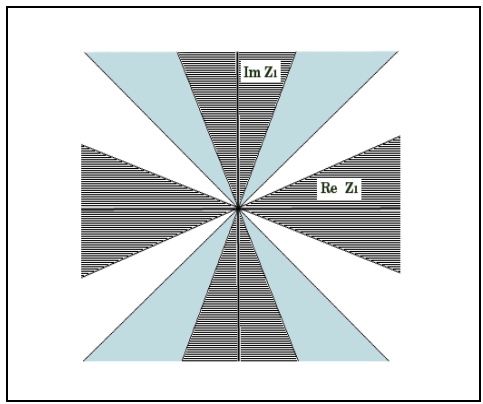}}
  \caption{(Color online): Stoke wedges: Dark portions denote the Stoke wedges where the wave function
  (\ref{phi}) is normalizable. Blue portions denote the Stoke wedges where the wave function
  (\ref{phi}) is normalizable when $\tilde{a}=0$ and $\tilde{b}<0$.  }
  \label{figure}
\end{center}
\end{figure}

\subsection{Rotationally invariant system}

For the choices $F^+_i=z^+_ig(r_i)$, $F^-_i=z^-_ig(r_i)$ and $V=V(\{r_i\})$
the Hamiltonian of Eq. (\ref{quantumH}) becomes
\bea
\hat{H} &=& \sum_{j=1}^m\left[\left(-i\partial_{z_j^+}+\frac{\gamma}{2}z^-_jg\right)^2 
-\left(-i\partial_{z_j^-}-\frac{\gamma}{2}z^+_jg\right)^2 \right]
+V(\{r_i\}),
\label{quantumH1}
\eea
where $r_i^2=(z_i^+)^2- (z_i^-)^2$. The pseudo-Euclidean angular momentum
operators,
\bea
\hat{L}_i=2\left(z^+_iP_{z^-_i}+z^-_iP_{z^+_i}\right),\ \ i=1,2,\dots m,
\eea
are integrals of motion and satisfy the following commutation relations
\bea
[\hat{H}, \hat{L}_i]=0,\ \ \ \ \ \  [\hat{L}_i, \hat{L}_j]=0,  \ i=1,2,\dots, m.
\label{2com}
\eea
The existence of $m+1$ integrals of motion implies that the system is at least
partially integrable and for $N=2,m=1$, the system is completely integrable.
 
\noindent An imaginary scale transformation of the form,
\bea
P_{z_i^-}\rightarrow i\tilde{P}_{z_i^-},\ z_i^-\rightarrow -i\tilde{z}_i^-,
P_{z_i^+}\rightarrow \tilde{P}_{z_i^+},\ z_i^+\rightarrow \tilde{z}_i^+,
\eea
may be performed to define the eigenvalue problem on a Euclidean plane.
In particular, the Hamiltonian of Eq. (\ref{quantumH1}) may be rewritten as,
\bea
\hat{H}&=&{\cal H}+i\gamma \cal{L},\nonumber\\
{\cal H}&=&\sum_{i=1}^m \left(\tilde{P}^2_{z_i^+}+\tilde{P}^2_{z_i^-}\right)
+ V_{eff}, \ \ {\cal L}= \frac{1}{2} \sum_{i=1}^m g(\tilde{r}_i) 
\tilde{L}_i,
\label{polar}
\eea
\noindent where the effective potential $V_{eff}$, the Euclidean angular
momentum operators $\tilde{L}_i$ and the Euclidean radial variables
$\tilde{r}_i$ are defined through the following relations:
\bea
V_{eff}\equiv V(\{\tilde{r}_i\}) - \sum_{i=1}^m
\frac{\gamma^2}{4}\tilde{r}_i^2g^2(\tilde{r}_i),\ \
\tilde{L}_i \equiv 2 \left(\tilde{z}^+_i\tilde{P}_{z_i^-}-
\tilde{z}^-_i\tilde{P}_{z_i^+}\right),\ \ 
\tilde{r}_i^2 \equiv(\tilde{z}_i^+)^2+(\tilde{z}_i^-)^2.
\eea
The quantum problem, after the imaginary scale transformation, is redefined in
a Hilbert space in which the operators ${\cal H}$ and ${\cal L}$ are
Hermitian \cite{ben-man}. Consequently, for a suitable choice of the
potential $V$, entirely real spectra with normalizable eigenfunctions may be
found for ${\cal H}$. The commutator of $\cal{H}$ and ${\cal L}$,
\bea
\left[{\cal H},{\cal L}\right]=-\frac{i}{2}\left\{\left(\tilde{P}_{z_i^+}\frac{\partial g}{\partial \tilde{z}_i^+}+
\frac{\partial g}{\partial \tilde{z}_i^+}\tilde{P}_{z_i^+}\right)+\left(\tilde{P}_{z_i^-}\frac{\partial g}{\partial \tilde{z}_i^-}+
\frac{\partial g}{\partial \tilde{z}_i^-}\tilde{P}_{z_i^-}\right)\right\}\tilde{L}_i,
\eea
vanishes only if gain/loss coefficient $g(\tilde{r}_i)$ is constant.
The Hamiltonian $\hat{H}$ can not admit entirely real spectra for constant
gain/loss coefficient, since the simultaneous eigenstates of ${\cal{H}}$ and
${\cal{L}}$ also diagonalize $\hat{H}$ and the corresponding energy eigenvalue
$\hat{E}$ contains an additive term of the form $i \gamma l$, where $l$ is the
eigenvalue of ${\cal{L}}$ \cite{ds-pkg1}. However, for space-dependent
gain/loss co-effiecient, entirely real spectra for $\hat{H}$ with normalizable
eigenfunctions are not ruled out completely. For example, the Hamiltonian
$\hat{H}$ in Eq. (\ref{polar}) for $N=2$ takes the following form in polar
coordinates $(\tilde{r}, \tilde{\theta})$ on the Euclidean plane,
\bea
\hat{H}=-\left(\frac{\partial^2}{\partial \tilde{r}^2}+\frac{1}{\tilde{r}}\frac{\partial}{\partial \tilde{r}}+
\frac{1}{\tilde{r}^2}\frac{\partial^2}{\partial \tilde{\theta}^2}\right)
 -\frac{\gamma^2}{4}g^2(\tilde{r})\tilde{r}^2+\gamma g(\tilde{r})\frac{\partial}{\partial \tilde{\theta}}+V(\tilde{r}).
\eea
The separation  of variables may be achieved by choosing
$\psi=\exp{(il\tilde{\theta})}\phi(\tilde{r})$ for which the stationery
eigenvalue equation with energy $E$ has the following form:
\bea
-\frac{\partial^2\phi}{\partial \tilde{r}^2}-\frac{1}{\tilde{r}}\frac{\partial\phi}{\partial \tilde{r}}+
\frac{l^2}{\tilde{r}^2}\phi+V_{\tilde{r}}\phi
=E\phi,\ \  V_{\tilde{r}}=V(\tilde{r})-\frac{\gamma^2}{4}g^2(\tilde{r})\tilde{r}^2+il\gamma g(\tilde{r}).
\label{sepsol}
\eea
It should be noted that the radial potential $V_{\tilde{r}}$ becomes complex due
to the presence of the $il\gamma g$ term. It may be recalled that within the
context of ${\cal{PT}}$ symmetric and/or pseudo-hermitian quantum system
complex potential may admit entirely real spectra with normalizable
eigenfunctions \cite{cmb,ali}. Thus, for space-dependent gain/loss coefficients, the
possibility of having a consistent quantum system with entirely real spectra
exists. However, even for the linear dependence of $g$ on $\tilde{r}$,
$V_{\tilde{r}}$ admits a quartic term. This makes the search for solvable models
nontrivial and is left for future investigations.

 \section{Summary and discussions}
 
The Hamiltonian formulation of a generic many-body system with space dependent
balanced loss and gain coefficient has been presented. It has been shown that
the balancing of loss and gain necessarily occurs in a pair-wise fashion. One
important aspect of this construction is that the use of an appropriate
orthogonal transformation allows the Hamiltonian to be interpreted as a
many-particle system in the background of a pseudo-Euclidean metric and
subjected to an analogous inhomogeneous magnetic field with a functional
form that is identical with space-dependent loss/gain co-efficient. The gauge
transformations of the analogous vector field correspond to various
Lagrangian  differing from each other by a total time-derivative term.

Discussions have been made on the choice of the potential which produces
unidirectional coupling between the system and the bath. In particular, the
dissipative dynamics of a system is independent of the dynamics of bath degrees
of freedom, while the converse is not true. Such a formulation has the
advantage that the techniques associated with Hamiltonian formulation, like
canonical perturbation theory, canonical quantization, KAM theory, geometric
mechanics etc. may be used to investigate purely dissipative dynamics of a
system. The examples presented are Hamiltonian corresponding to rational
as well as trigonometric dissipative Calogero-Sutherland models with various
root systems and dissipative Toda systems.

The equations of motion resulting from the Hamiltonian are coupled
Li$\acute{e}$nard type differential equations with balanced loss and gain.
Special emphasize have been given to investigate the integrability and exact
solvability of the system. Two specific classes of models with $N=2m$ number
of particles admitting translational or rotational symmetry have been
investigated in some detail. A total number of $m+1$ integrals of motion have
been constructed for the both types of systems, which are in involution,
implying that the system is partially integrable for $N>2$ and is completely
integrable for $N=2$. The space-dependent gain-loss co-efficients make the
equations of motion nonlinear irrespective of the specific form of the
potential. This makes the search for exact solutions nontrivial. Nevertheless,
for both the cases, exact solutions are obtained for a few specific choices of
the potentials and space-dependent gain/loss co-efficients.

The quantization of the system with space dependent balanced loss and gain has
been carried out. The $m+1$ number of quantum integrals of motion are
constructed for a system of $2m$ particles with translational or rotational
symmetry. It appears that solving the complete eigenvalue problem analytically
is a nontrivial task, even for potentials like harmonic oscillator, Coulomb,
etc. due to the space-dependent balanced loss/gain terms. For example, a choice
of the gain/loos co-efficient depending linearly on one of the co-ordinates
produces a quartic term in the same co-ordinate in the eigenvalue equation.
A class of quasi-exactly solvable models with translational symmetry has been
presented in this article with a discussion on the normalizability of the
wave-function in appropriate Stoke wedges. For the case of rotationally
symmetric system, attempts to find solvable or quasi-exactly solvable models
admitting bound states have not produced any positive result. However, unlike
the case of constant loss/gain coefficients\cite{pkg-ds}, the possibility of
rotationally symmetric system with space-dependent balanced loss/gain
coefficients admitting bound states is not completely ruled out.
 
 \section{Acknowledgments}
 {\bf DS} acknowledges a research
fellowship from CSIR.

\section{Appendix-A: Gauge transformations and equivalent Lagrangian}

In this appendix the Lagrangian corresponding to the Hamiltonian in  Eq. (\ref{hz}) that is relevant in the present discussions
is presented. As has been mentioned in section-2 that
the Hamiltonian in Eq. (\ref{hz}) may be interpreted as describing a many-particle system subjected to an external inhomogeneous magnetic 
field and the gauge transformations of the vector potential corresponding to external magnetic field produce
Lagrangian that differs from (\ref{Lz}) by a total time derivative term and is equivalent to each other 
in the sense that they lead to the same equations of motion.  The Lagrangian presented in this appendix is 
particularly useful when the system admits certain symmetries. For example, the following Lagrangian may be considered 

\bea
L_{1}=\sum_{i=1}^m\left[\frac{1}{4}\left\{(\dot{z}_i^+)^2-(\dot{z}_i^-)^2\right\}-\frac{\gamma}{2}\dot{z}_i^+\int Q_i(z_i^+,z_i^-)dz_i^-\right]-V(\{z_i^+,z_i^-\}).
\label{L1}
\eea
For translationally symmetric system with $Q_i=Q_i(z_i^-)$ and $V=V(z_i^-)$, the coordinates $z_i^+$ become
cyclic which leads to $m$ conserved quantities. In this case the Routhian of the system  has the 
following form

\bea
R_1=\sum_{i=1}^m\left[\left(P_{z_i^+}+\frac{\gamma}{2}\int Q_i(z_i^-)dz_i^-\right)^2+\frac{1}{4}(\dot{z}_i^-)^2\right]+V(\{z_i^-\}).
\eea
Another Lagrangian corresponding to the Hamiltonian in  Eq. (\ref{hz}) may be presented as follows

\bea
L_{2}=\sum_{i=1}^m\left[\frac{1}{4}\left\{(\dot{z}_i^+)^2-(\dot{z}_i^-)^2\right\}+\frac{\gamma}{2}\dot{z}_i^-\int Q_i(z_i^+,z_i^-)dz_i^+\right]-V(\{z_i^+,z_i^-\}).
\label{L2}
\eea
This Lagrangian is specially convenient to use when the system yields a symmetry such that 
 the coordinates $z_i^-$ become cyclic with $Q_i$ and $V$ are respectively given by $Q_i=Q_i(z_i^+)$ and $V=V(z_i^+)$.
 In this case the Routhian corresponding to the Lagrangian (\ref{L2}) takes the following form,
 
 \bea
 R_2=\sum_{i=1}^m-\left[\frac{1}{4}(\dot{z}_i^+)^2+\left(P_{z_i^-}-\frac{\gamma}{2}\int Q_i(z_i^+)dz_i^+\right)^2\right]+V(\{z_i^+\}).
 \eea
The Hamiltonian equations corresponding to the Routhains $R_1$ and $R_2$ give the constants of motion and the Lagrangian
equations give the equations of motion corresponding to the non-cyclic coordinates in the respective cases.

\end{document}